\documentstyle[aps,prl,preprint,floats,epsf]{revtex}

\begin{document}
\draft
\tighten

\preprint{\vbox{\hbox{\hfil CLNS 97/1493}
                        \hbox{\hfil CLEO 97-14}
}}

\title{A measurement of the total cross section for
$e^+e^-\to$hadrons at $\sqrt{s}$=10.52 GeV}
\maketitle

\begin{abstract}
Using the CLEO detector at the Cornell Electron Storage Ring,
we have made a measurement of 
$R\equiv{\sigma(e^+e^-\to hadrons)\over\sigma(e^+e^-\to\mu^+\mu^-)}
=3.56\pm0.01\pm0.07$ at
$\sqrt{s}$=10.52 GeV.
This implies a value for the strong coupling constant 
of $\alpha_s(10.52~{\rm GeV})=0.20\pm0.01\pm0.06$, or 
$\alpha_s(M_Z)=0.13\pm0.005\pm0.03$.
\end{abstract}
{\it\centerline{Submitted to Physical Review D}}

\newpage
\begin{center}
R.~Ammar,$^{1}$ P.~Baringer,$^{1}$ A.~Bean,$^{1}$
D.~Besson,$^{1}$ D.~Coppage,$^{1}$ C.~Darling,$^{1}$
R.~Davis,$^{1}$ N.~Hancock,$^{1}$ S.~Kotov,$^{1}$
I.~Kravchenko,$^{1}$
% ...
\end{center}
 
\small
\begin{center}
$^{1}${University of Kansas, Lawrence, Kansas 66045}\\
% ...
$^{12}${Virginia Polytechnic Institute and State University,
Blacksburg, Virginia 24061}\\
$^{13}${Wayne State University, Detroit, Michigan 48202}\\
% ...
\end{center}
 
\newpage

\section{Introduction}
The measurement of the hadronic production cross section
in $e^+e^-$ annihilation is perhaps the most fundamental
experimentally accessible quantity in quantum chromodynamics (QCD)
due to its insensitivity to the fragmentation process.
The measured hadronic cross-section is generally expressed in
terms of its ratio $R$ to the point cross-section for $\mu^+\mu^-$
production.
In QCD, $R$ is directly
proportional to the number of colors, depends on quark charges,
and varies with energy, both discreetly as quark mass thresholds are crossed,
and gradually as the strong coupling constant 
$\alpha_s$ ``runs". So, historically, $R$ measurements have been
valuable in verifying quark thresholds, charges, color-counting,
and the existence of the gluon. 

The theoretical 
prediction for $R$ is 
\begin{equation}
R=R_{(0)}(1+\alpha_s/\pi+C_2(\alpha_s/\pi)^2+C_3(\alpha_s/\pi)^3). \label{eq:R}
\end{equation}
\noindent
A calculation appropriate at LEP energies
obtained
$C_2=1.411$ and $C_3=-12.68$\cite{ruski91} for five
active flavors, in the limit of massless quarks.
A recent calculation, applicable to the $\Upsilon$ mass region 
(four active flavors), has
included corrections due to 
the effects of quark masses and QED radiation to obtain
$C_2$=1.5245 and $C_3$=--11.52
at $\sqrt{s}$=10 GeV\cite{Teubner96}. 
The effect of including these
additional corrections is a difference of approximately 0.3\% in the 
prediction for $R$ at this energy.
$R_{(0)}$ is the lowest-order prediction 
for this ratio, given by $R_{(0)}=N_c\Sigma_i q_i^2$, where $N_c$ is the
number of quark colors;
the sum runs over the kinematically allowed quark flavors.
Just below the $\Upsilon$(4S) resonance, where b\=b production
is kinematically forbidden,
the lowest-order prediction is therefore obtained by summing over
$udcs$ quarks, yielding $R_{(0)}$=10/3. 
An experimental measurement of $R$ can
therefore
be used to deduce a value for $\alpha_s$.
In this Article we present a measurement of $R$ 
using the CLEO detector operating at the Cornell Electron 
Storage Ring (CESR) at a center-of-mass energy $\sqrt{s}$=10.52 GeV.

\section{Apparatus and Event Selection}
The CLEO~II detector is a general purpose solenoidal magnet spectrometer and
calorimeter\cite{r:CLEO-II}. 
The detector was
designed to trigger efficiently on
two-photon, tau-pair, and hadronic events.
As a result, although hadronic 
event reconstruction efficiencies are high, lower-multiplicity
non-hadronic backgrounds require careful consideration in this analysis.
Good background rejection is afforded by the high precision electromagnetic
calorimetry and excellent charged particle tracking capabilities.
Charged particle momenta are measured with
three nested coaxial drift chambers with 6, 10, and 51 layers,
respectively.  These chambers fill the volume from $r$=3 cm to $r$=100 cm, 
where
$r$ is the radial coordinate relative to the beam ($z$) axis, and have
good efficiency for charged particle tracking for polar angles
$|cos\theta|<$0.94, with $\theta$ measured 
relative to the positron beam direction ($+{\hat z}$).  
This system achieves a momentum resolution of $(\delta p/p)^2 =
(0.0015p)^2 + (0.005)^2$, where $p$ is momentum in GeV/c. 
Pulse height measurements in the main drift chamber provide specific
ionization resolution
of 6.5\% for Bhabha events, giving good $K/\pi$ separation for tracks with
momenta up to 700 MeV/c and approximately 2 standard
deviation resolution in the relativistic
rise region. Outside the central tracking chambers are plastic
scintillation counters that are used as fast elements in the trigger system
and also provide particle identification information from time-of-flight
measurements. 
Beyond the time-of-flight system is the electromagnetic calorimeter,
consisting of 7800 thallium-doped cesium
iodide crystals.  The central ``barrel'' region
of the calorimeter covers about 75\% of the solid angle and has an 
energy resolution of about 4\% at 100 MeV and 1.2\% at 5 GeV. Two endcap
regions of the crystal calorimeter extend solid angle coverage to about 98\%
of $4\pi$, although with somewhat worse energy resolution than the 
barrel region.
The tracking system, time-of-flight counters, and calorimeter
are all contained 
within a 1.5 T superconducting coil. 

\begin{figure}[phtb] \vspace{10 cm}
\includegraphics{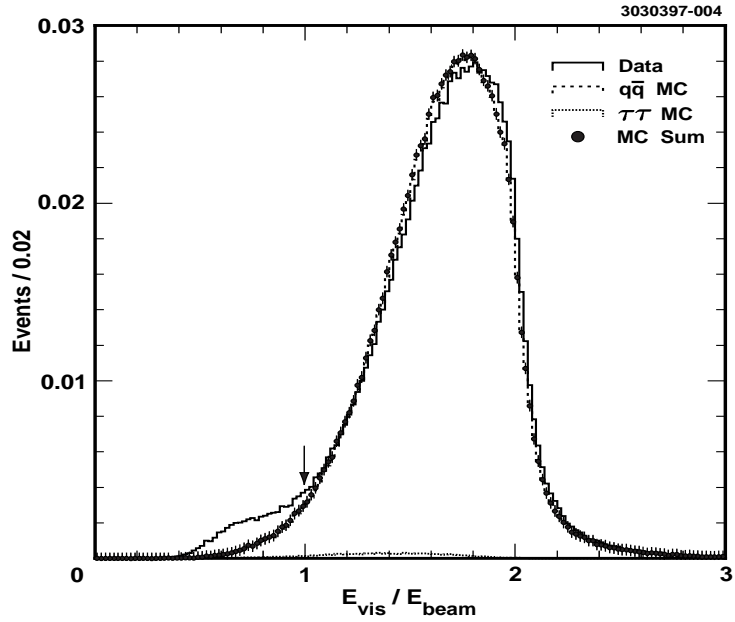}
\caption{\label{fig:3030397-004.ps}Normalized
visible energy distribution 
for data
(solid), q\=q Monte Carlo (dashed), and $\tau\tau$ Monte Carlo (dotted).
Sum of q\=q plus $\tau\tau$ Monte Carlo is shown as crosses.}
\end{figure}

\begin{figure}[phtb] \vspace{10 cm}
\includegraphics{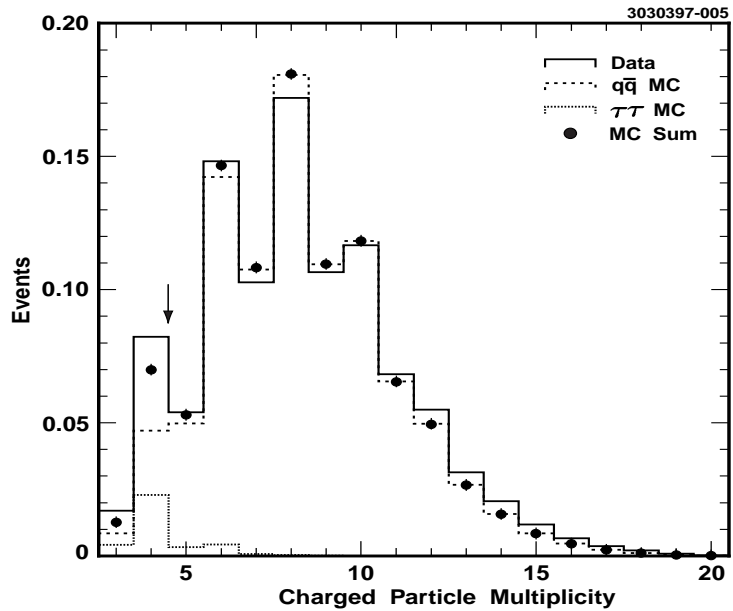}
\caption{\label{fig:3030397-005.ps}
Normalized charged 
multiplicity distribution 
for data vs. Monte Carlo.}
\end{figure}

\begin{figure}[phtb] \vspace{10 cm}
\includegraphics{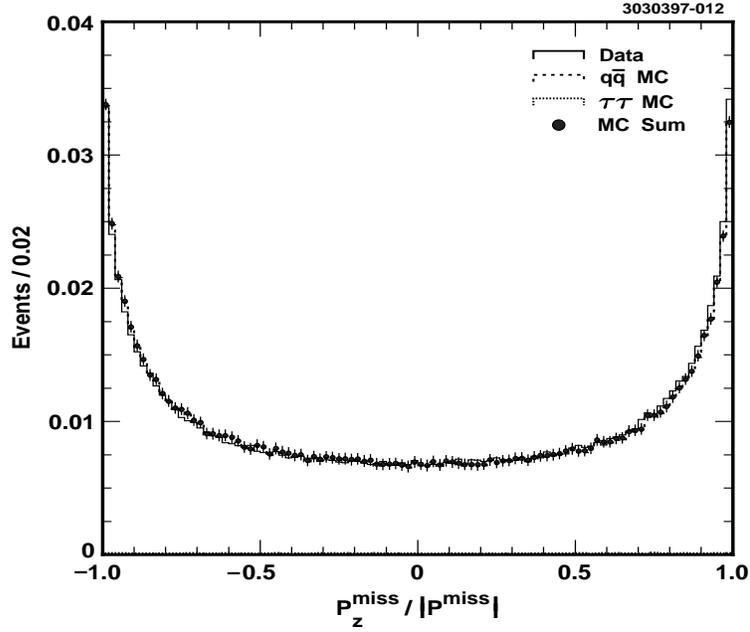}
\caption{\label{fig:3030397-012.ps}Distribution of
z-component (i.e., direction cosine) 
of missing momentum $P_z^{miss}/|P^{miss}|$ 
for data vs. Monte Carlo (this variable is
not cut on).}
\end{figure}

\begin{figure}[phtb] \vspace{10 cm}
\includegraphics{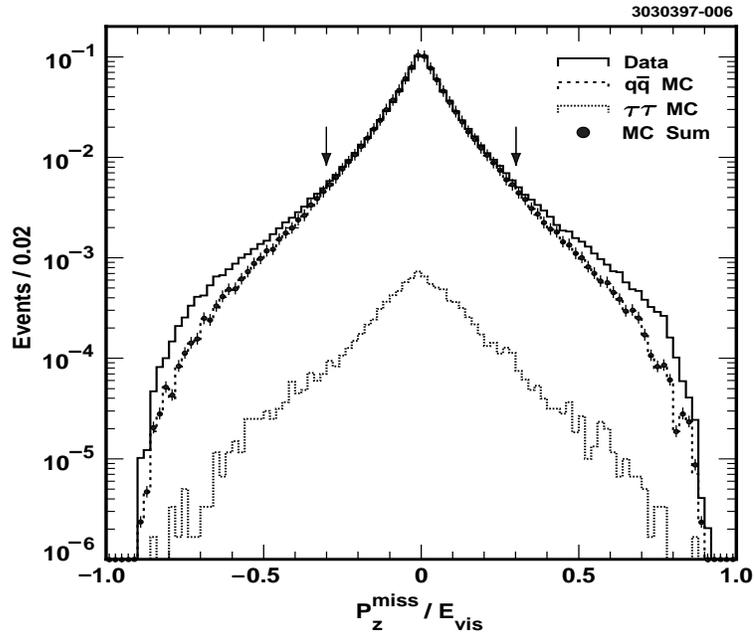}
\caption{\label{fig:3030397-006.ps}Ratio of $P_z^{miss}/E_{visible}$
for data vs. Monte Carlo. Two-photon collisions, and beam-gas interactions
tend to populate the regions away from zero and towards
$\pm$1 in this plot.}
\end{figure}

\begin{figure}[phtb] \vspace{10 cm}
\includegraphics{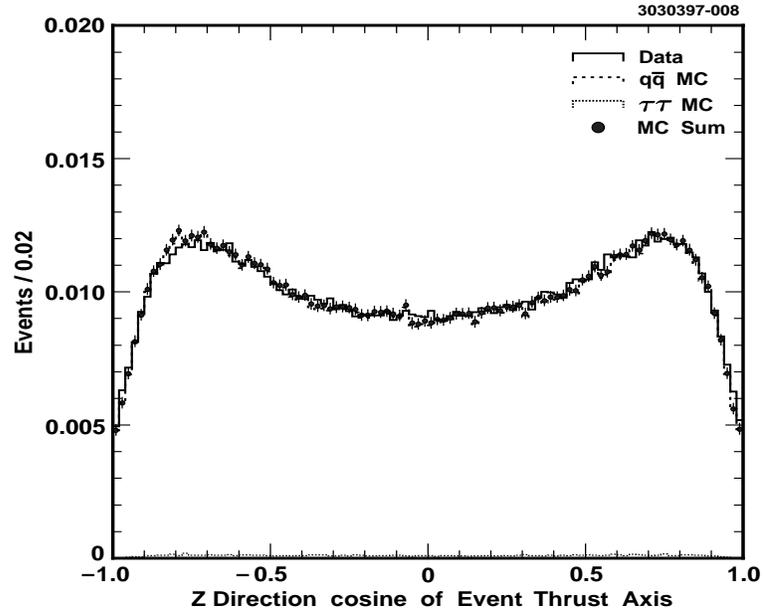}
\caption{\label{fig:3030397-008.ps}Z-component of thrust axis for data vs. 
Monte Carlo (this variable is not cut on).}
\end{figure}

\begin{figure}[phtb] \vspace{10 cm}
\includegraphics{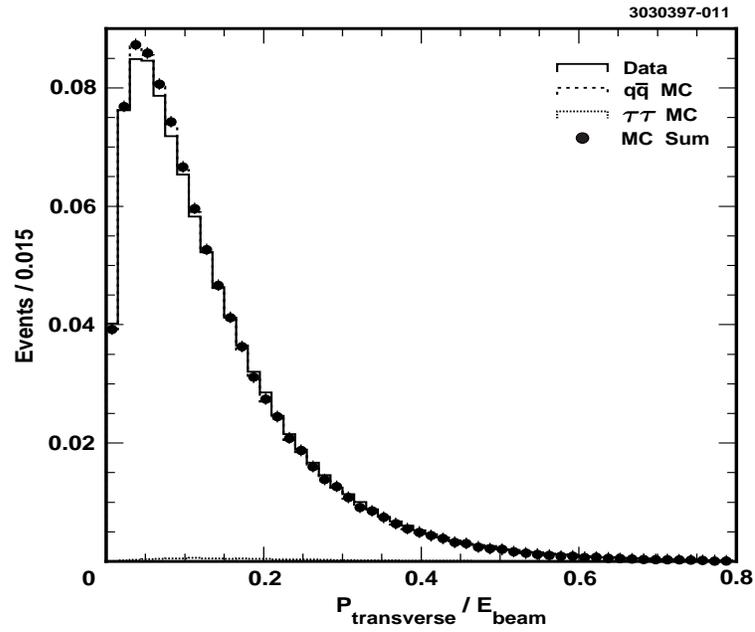}
\caption{\label{fig:3030397-011.ps}Ratio of transverse momentum relative to 
visible energy (this variable is not cut on).}
\end{figure}

\begin{figure}[phtb] \vspace{10 cm}
\includegraphics{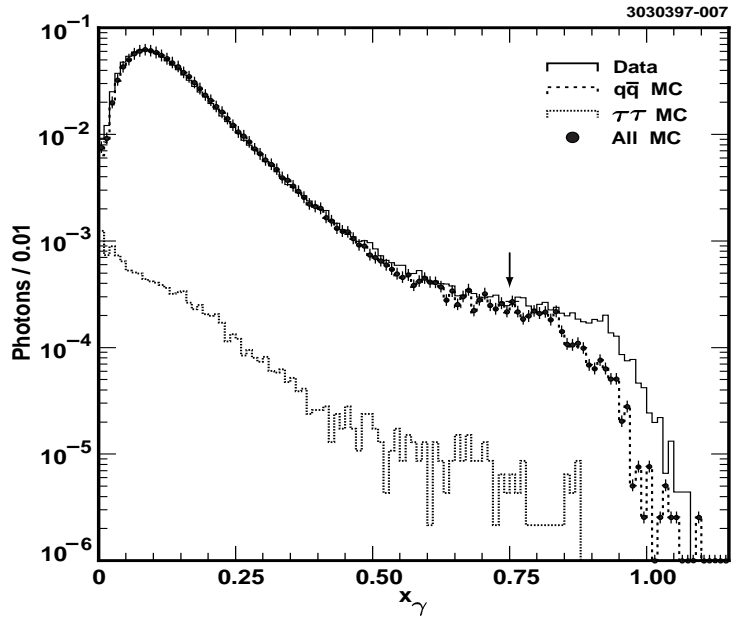}
\caption{\label{fig:3030397-007.ps}Comparison 
of data vs. Monte Carlo spectrum of
most energetic photon observed in event.
}
\end{figure}

To suppress $\tau\tau$, $\gamma\gamma$, low-multiplicity
QED, and other backgrounds while maintaining relatively high 
q\=q event reconstruction efficiency,
the following 
requirements are imposed to define our hadronic event sample:

\begin{enumerate}
\item At least 5 detected, good quality, 
charged tracks ($\tt N_{chrg}$) $\ge$5.
\item The total visible energy ${\tt E_{vis}}$ (=$E_{chrg}+E_{neutral}$) 
should be
greater than the single beam
energy: ${\tt E_{vis}>E_{beam}}$.
\item The z-component of the 
missing momentum must satisfy: ${\tt |P_z^{miss}|\over E_{vis}}<$0.3.
\item To suppress events originating as collisions of $e^\pm$ beam particles
with gas or the vacuum chamber walls, we require that the
reconstructed event vertex (defined as $z_{vrtx}$)
be within 5.5~cm in $z$ (${\hat z}$ defined above as the 
$e^+$ beam direction) and
2~cm in cylindrical radius of the nominal interaction point.
\item In addition to these primary requirements, 
additional criteria are imposed to 
remove backgrounds remaining at the $\sim$1\% level, as well
as to suppress events with hard initial state radiation, for which
theoretical uncertainties are large. These are:
\begin{enumerate}
\item No more than
two identified electrons are in the event.
\item The ratio $R_2$ of the second to the zeroth
Fox-Wolfram moments\cite{FoxWolf} for the event should
satisfy $R_2<$0.9.
\item The ratio of calorimeter energy contained in showers that match to
charged particles 
relative 
to the scalar sum of the momenta of all the charged particles in the
drift chamber ($E\over p$)
must be less than 0.9.
\item The most energetic 
photon candidate detected in the event must have
a measured energy less than 0.75 of
the beam energy ($x_\gamma\equiv{E_\gamma^{max}\over E_{beam}}<$0.75).
This requirement reduces the uncertainty from radiative corrections.
\end{enumerate}
\end{enumerate}
Figures \ref{fig:3030397-004.ps}--\ref{fig:3030397-007.ps}
show comparisons between our candidate hadronic
event sample in data 
and the Monte 
Carlo sample (arrows indicate our cut values). Simulated hadronic events
are produced using the
JETSET 7.3 q\=q event generator\cite{r:LUND} 
run through a full GEANT-based\cite{r:GEANT} 
detector simulation. 
Tau-pair events 
use the KORALB\cite{KORALB} 
event
generator in conjunction with the same detector simulation.
In all of these comparison plots, 
both the data and the `Monte Carlo sum' have been normalized to 
unit area in the `good' acceptance region.
We also use this Monte Carlo event sample to
determine the efficiency for q\=q events to pass our hadronic
event selection requirements. 
We note that,
according to the Monte Carlo simulation, the trigger
inefficiency with the default event selection 
criteria is less than 0.1\%.
This has been checked with the data by counting how
many hadronic events triggered only a minimum bias, pre-scaled
trigger line.

\section{Backgrounds}
After imposition of the above event selection criteria, 
we are left with a sample of $4.00\times 10^6$
candidate hadronic events. 
Small
backgrounds still remain.
These are enumerated as follows:
\begin{enumerate}
\item Backgrounds from 
$e^+e^-\to\tau^+\tau^-(\gamma)$ events are subtracted statistically
using a large Monte Carlo sample of KORALB tau-pair events.
These events
comprise $(1.3\pm0.1)$\% (statistical error only) 
of the sample passing the above event selection criteria.
\item Backgrounds from the narrow $\Upsilon$ resonances 
can be explicitly determined from data using
$e^+e^-\to\gamma\Upsilon(3S/2S)$; $\Upsilon(3S/2S)\to\pi^+\pi^-\Upsilon(1S)$;
$\Upsilon(1S)\to l^+l^-$ events. 
These events are distinctive by their characteristic topology of two 
low-momentum pions accompanied by two very 
high momentum, back-to-back leptons;
the photon
generally escapes undetected along the beam axis.
As shown in 
Figure \ref{fig:3030397-010.ps},
\begin{figure}[phtb] \vspace{9 cm}
\includegraphics{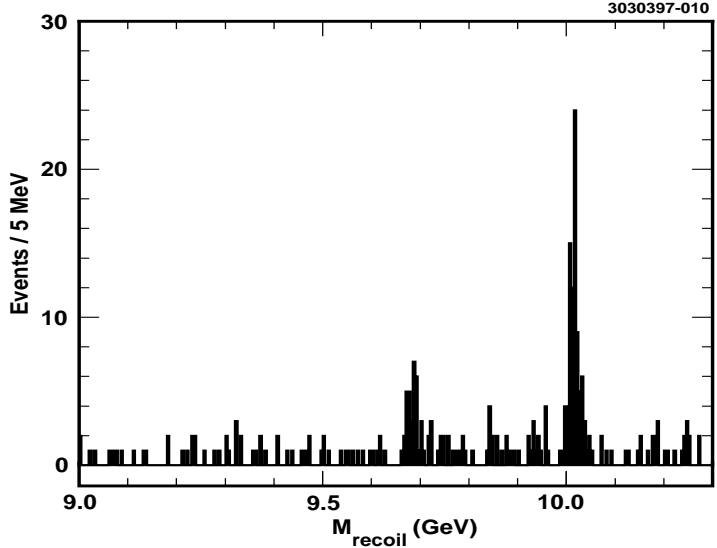}
\caption{\label{fig:3030397-010.ps}
Mass recoiling against two charged particles,
assumed to be pions, in events consistent with the kinematics for:
$e^+e^-\to\gamma\Upsilon(3S/2S); 
\Upsilon(3S/2S)\to\Upsilon(1S)\pi^+\pi^-; \Upsilon(1S)\to l^+l^-$. Two peaks
are evident; the leftmost peak corresponds to 
$\Upsilon$(3S)$\to\pi^+\pi^-\Upsilon$(1S) transitions, the rightmost
peak corresponds to 
$\Upsilon$(2S)$\to\pi^+\pi^-\Upsilon$(1S) transitions. The calculated
recoil mass differs from the true $\Upsilon$(1S) mass due to 
our neglecting 
the (undetected) radiated
photon in the recoil mass calculation.}
\end{figure}
we observe these events 
as distinct peaks in the mass distribution recoiling
against two low-momentum pions in events also containing two 
high-energy muons.
(The recoil mass is calculated from: $M_{recoil}=\sqrt{(2E_{beam} - E_{\pi_1}
- E_{\pi_2})^2-({\vec p_{\pi_1}}+{\vec p_{\pi_2}})^2}$, and therefore
neglects the four-momentum of the initial state radiation photon.)
Backgrounds from QED processes ($e^+e^-\to\gamma\ell^+\ell^-$; 
$\gamma\to e^+e^-$) can be suppressed by requiring that the candidate
dipion system not be colinear with either of the final state leptons.
Knowing the branching fractions\cite{PDG96} for 
$\Upsilon$(2S)$\to\pi\pi\Upsilon$(1S) 
($18.5\pm0.8$\%)
and $\Upsilon$(3S)$\to\pi\pi\Upsilon$(1S) ($4.5\pm0.2$\%), 
the leptonic branching fraction for the $\Upsilon$(1S) ($2.5\pm0.1)$\%,
and the reconstruction efficiency for such events
($\sim$0.7), we can determine the
contribution to the observed hadronic cross-section from the 
$\Upsilon$(2S) and $\Upsilon$(3S) resonances directly, by
simply measuring the event yields in the peaks shown in 
Figure \ref{fig:3030397-010.ps}, and correcting by branching fractions
and efficiency. 

For the
contribution from the $\Upsilon$(1S) resonance, we proceed by
assuming that the photon spectrum varies as: 
${dN\over dE_\gamma}\sim {1\over E_\gamma}$,
and that the production of
a given Upsilon resonance is proportional to its dielectron width
$\Gamma_{ee}$.
This gives a fairly simple prediction for the ratios expected for 
a given Upsilon resonance, since $E_\gamma\sim(10.52-M_\Upsilon$) GeV.
We would expect that the production cross-section for $\Upsilon\gamma$ in
$e^+e^-$ annihilation therefore varies as:
$\Gamma(e^+e^-\to\Upsilon\gamma)\propto{\Gamma_{ee}^\Upsilon\over E_\gamma}.$
This allows us to infer an expected production cross-section for
$\gamma\Upsilon(1S)$ based on our measurements for
$\gamma\Upsilon(2S)$ and $\gamma\Upsilon(3S)$ production.
We compare our extrapolated
cross-section for $e^+e^-\to\gamma\Upsilon$(1S) through
this procedure with theory in order to estimate the magnitude of
this correction.
Combining our data with the theoretical predictions of 
Teubner et al.\cite{Teubner96},
we determine that the 
$\gamma\Upsilon$(1S), $\gamma\Upsilon$(2S), 
and $\gamma\Upsilon$(3S) events comprise
(1.8$\pm$0.6)\% of the observed hadronic cross-section,
where the error includes the uncertainties in the Upsilon decay
branching fractions and detection efficiencies as well as the
deviations between the theoretical and measured values.
\item Two-photon collisions, which produce hadrons in the
final state via $e^+e^-\to e^+e^-\gamma\gamma\to e^+e^-$+hadrons, are
determined by running final-state specific $\gamma\gamma$ collision Monte
Carlo events, and also by determining the magnitude of possible excesses in
the $E_{visible}~vs.~P_{transverse}$ plane for data over q\=q 
Monte Carlo. These
are determined to comprise (0.8$\pm$0.4)\% of our total hadronic event sample. 
\item Beam-wall, beam-gas, and cosmic ray events are expected to
have a flat event vertex distribution in the 
interval $|z_{vrtx}|<$10 cm;
such events are subtracted using a
sample having a vertex in the interval $5.5cm<|z_{vrtx}|<10cm$,
and extrapolating into the `good' acceptance region
($|z_{vrtx}|<$5.5 cm). These backgrounds are
determined to comprise $\sim(0.2\pm0.1)$\% of our hadronic sample.
\item Remaining QED backgrounds producing 2 or more electrons or muons in
the final state are assessed 
using a
high-statistics sample of Monte Carlo events (to $3^{rd}$ 
order in $\alpha_{QED}$),
and found to be $\le0.1$\% of the sample passing the above hadronic event 
selection requirements.
\end{enumerate}
Summing these estimates results in a net background
fraction $f=(4.1\pm 0.7)\%$. We note that, as this error is assessed partly
by examining the difference between Monte Carlo hadronic event simulations
and our data, this error also includes Monte Carlo modeling errors.

\section{Efficiencies and Radiative Corrections}

The computation of $R$ is peformed with
\begin{equation}
R={N_{had}(1-f)\over{\cal L}\epsilon_{had}(1+\delta)\sigma_{\mu\mu}^0},
\label{eq:Rexpt}
\end{equation}
\noindent
where $N_{had}$ is the number of events classified
as hadronic, $f$ is the
fraction of selected events attributable to all background
processes, 
$\epsilon_{had}$ is
the efficiency for triggering and selection of events,
$\delta$ is the fractional increase in hadronic cross section
due to electromagnetic radiative corrections to
that cross section, $\sigma_{\mu\mu}^0$
is the point cross section for muon pair production 
($86.86 nb/E_{cm}^2 (GeV^2)$),
and ${\cal L}$ 
is the measured integrated luminosity. 
The luminosity is determined from wide angle
$e^+e^-$, $\gamma\gamma$,
and $\mu^+\mu^-$ final states and is known to $\pm$1\%\cite{NIMLUM}.
For the data analyzed here, the integrated luminosity
${\cal L}$ is equal to $(1.521\pm 0.015)$ fb$^{-1}$.

\begin{figure}[phtb] \vspace{10 cm}
\includegraphics{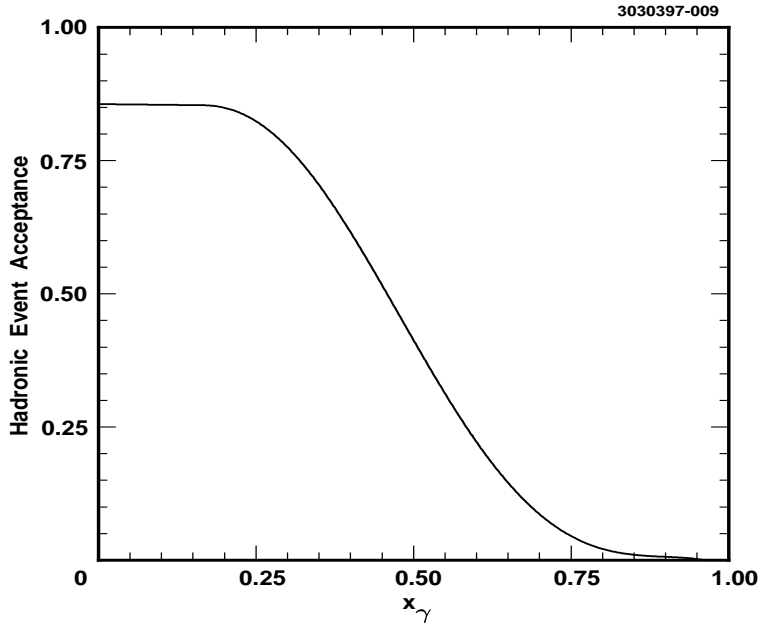}
\caption{\label{fig:3030397-009.ps}Efficiency for 
reconstructing $e^+e^-\to\gamma q{\overline q}$ event, 
as a function of the 
scaled photon momentum $x_\gamma \equiv E_\gamma/E_{beam}$.}
\end{figure}

To calculate $R$, we must therefore evaluate Eqn. \ref{eq:Rexpt}.
If the initial state radiation corrections were known precisely,
we would be able to calculate the 
denominator term $\epsilon(1+\delta)$ with
very good precision. However, since the uncertainties become
very large as the center-of-mass energy approaches c\=c threshold 
($\sqrt{s}\sim$4
GeV), the preferred procedure is to choose some explicit
cut-off in initial state radiation (ISR)
photon energy that makes us as insensitive as possible to the corrections
in this high ISR photon energy/low hadronic
recoil mass region. 
We therefore purposely design our selection criteria
so that our efficiency for events with
highly energetic
ISR photons approaches zero. 
By choosing cuts that drive $\epsilon$ to zero beyond some kinematic point,
we ensure that the product $\epsilon\times(1+\delta)$ is insensitive to
whatever value of $\delta$ may be prescribed by theory beyond our cut.
Thus,
although there is a large uncertainty
in the magnitude of the initial state radiation correction for large
values of radiated photon momentum, we have minimized our sensitivity to this
theoretical uncertainty.
Figure \ref{fig:3030397-009.ps}
displays our efficiency for an $e^+e^-\to\gamma q{\overline q}$
event to pass our hadronic event criteria as a function of the scaled photon
energy $x_\gamma\equiv{E_\gamma\over E_{beam}}$. 
We note that for
$x_\gamma>$0.75 (corresponding to a $q{\overline q}$ recoil mass of
$M_{recoil}<5.25~GeV/c^2$), 
our integrated event-finding efficiency $\epsilon_{had}<$1\%. For 
$x_\gamma>$0.75, 
we have therefore minimized our
sensitivity to modeling uncertainties in
this kinematic regime -- increasing the 
initial state radiation contribution to this region results in a
compensating loss of overall efficiency such that the product of
$\epsilon(1+\delta)$ remains relatively constant. For our event
selection criteria, we thus select
the value of $\epsilon(1+\delta)(x_\gamma^{max}=0.75)=0.90\pm$0.01, where
the error reflects the systematic uncertainty in the radiative corrections.

After subtracting all backgrounds, dividing by
the total luminosity, and normalizing to the mu-pair point cross-section,
we obtain a value of $R=3.56\pm0.01$ (statistical error only).

\section{Systematic Errors and Consistency Checks}
We have checked our results in several ways. 
Backgrounds can be suppressed significantly by tightening the minimum
charged track multiplicity to $N_{chrg}\ge$7, 
albeit at a loss of $\sim$20\% in the overall
event-reconstruction efficiency. Imposition of such a cut leads to
only a
$-$0.4\% change in the calculated value of $R$.
Continuum data have been
collected over 17 distinct periods from 1990-1996, covering many different
trigger configurations and running conditions. We find a 0.3\% rms
variation 
between the various datasets used (the statistical error on R
within each
dataset is of order 0.1\%). We can check contributions 
due to the narrow Upsilon
resonances by calculating $R$ using a small amount ($5pb^{-1}$)
of continuum data taken just below the $\Upsilon$(2S) resonance, at
$E_{beam}$=4.995 GeV. We find that the value of $R$ calculated using the 
$\Upsilon$(2S)
continuum agrees with that calculated using the 
$\Upsilon$(4S) continuum to within
one statistical error ($1\sigma_{stat}$).
Systematic errors are summarized in Table \ref{tab:syserr}.

\begin{center}
\underline{Systematic error summary}
\begin{table}[hptb]
\begin{tabular}{|c|c|}
Source & Error \\
$\epsilon\times(1+\delta)$ & 1\% \\
${\cal L}$ & 1\% \\
Background Uncertainty/Hadronic Event Modeling Uncertainty & 0.7\% \\
Dataset-to-dataset variation & 0.3\% \\ \hline
Total & 1.8\% \\
\end{tabular}
\caption{\label{tab:syserr}Systematic errors in $R$ analysis.}
\end{table}
\end{center}

\section{Extraction of $\alpha_s$}
Using the expansion for $R$ in powers of $\alpha_s/\pi$ given
previously, with coefficients appropriate for this
center of mass energy\cite{Teubner96},
we can evaluate the strong coupling constant.
Using that
expression, our value for $R$ 
translates to $\alpha_s(10.52~GeV)=0.20\pm0.01\pm0.06$.

The strong coupling constant $\alpha_{\rm s}$ can be written
as a function of
the basic QCD parameter $\Lambda_{\overline{MS}}$, defined in the modified
minimal subtraction scheme~\cite{PDG96},
\begin{equation}
    \alpha_{\rm s}(\mu) = {4\pi \over b_0 x}
                          \Biggl(1 - {2b_1 \over b_0^2}{ln(x)\over x} +
	{4b_1^2\over b_0^4x^2}\times
                          ([ln(x) - \frac{1}{2}]^2 + {b_2b_0\over 8b_1^2}
	- \frac{5}{4})\Biggr) \label{eq:qcd3}
\end{equation}
\noindent
where 
$b_0=(11-2n_f/3)$, 
$\mu$ is the energy scale, in GeV, at which
$\alpha_s$ is being evaluated,
$b_1 = {51 - 19n_f \over 3}$, 
$b_2 = 2857 - {5033 n_f\over 9} + {325 n_f^2\over 27}$,
$x=ln(\mu^2/\Lambda_{\overline{MS}}^2)$,
and $n_f$ is the number of light quark flavors
which participate in the process.
To determine the value of $\alpha_s(90~GeV)$ 
implied by our measurement, we must evolve $\alpha_s$ across the
discontinuity in $\Lambda_{\overline{MS}}$ when the five-flavor
threshold is crossed from the four-flavor regime. 
We do so using the next-to-next-to leading order (NNLO)
prescription, as described in \cite{PDG96}:
a) We substitute
$\alpha_s$(10.52) into
Eqn. (\ref{eq:qcd3})
to determine a value for $\Lambda_{\overline{MS}}$ in the four-flavor
continuum (obtaining $\Lambda_{\overline{MS}}(udcs)$=498 MeV). 
b) With that value of $\Lambda_{\overline{MS}}$, we can now
again use Eqn. (\ref{eq:qcd3}) to determine the value of
$\alpha_s$ at the
five-flavor threshold when the $b$-quark pole mass (we use 
$m_{b,pole}$=4.7 GeV)
is crossed,
and then use that value of
$\alpha_s$, as well as $n_f=5$ in Eqn. (\ref{eq:qcd3})
to determine $\Lambda_{\overline{MS}}$ appropriate for the five-flavor
continuum.
c) Assuming that this value of
$\Lambda_{\overline{MS}}$ is
constant in the entire five-flavor energy region, we can
now evolve $\alpha_s$ up to the $Z$-pole, to obtain
$\alpha_s(M_Z)=0.13\pm0.005\pm 0.03$, in good
agreement with the world average $\alpha_s(M_Z)=0.118\pm0.003$\cite{PDG96}.

\section{Summary}
Near $\sqrt{s}$=10 GeV, $R$ has been measured by
many experiments, as shown in 
Table \ref{tab:Xscts}. 
The measurement of $R$ 
described here
is the most precise below the $Z^0$. Theoretical
uncertainties in QED radiative corrections (in the acceptance 
($\epsilon\times(1+\delta)$) and luminosity \cite{NIMLUM}) contribute about the
same amount to the systematic error as do backgrounds and efficiencies.
Substantially improving this measurement will require progress
on radiative corrections as well as on experimental techniques.
Our $R$ value is in good agreement with the previous world average,
including a recent determination by the 
MD-1 Collaboration\cite{MD1-96}. 
Our implied value of
$\alpha_s$ is in agreement with higher energy determinations of this
quantity.

\begin{table}[htb]
\begin{tabular}{ccc}
Experiment & $\sqrt{s}$(GeV) & $R$ \\ \hline
PLUTO\cite{PLUTO} & 9.4 & $3.67\pm0.23\pm0.29$ \\
DASPII\cite{DASPII} & 9.4 & $3.37\pm0.16\pm0.28$ \\
DESY-Heidelberg\cite{Heidelberg} & 9.4 & $3.80\pm0.27\pm0.42$ \\
LENA\cite{LENA} & 9.1-9.4 & $3.34\pm0.09\pm0.18$ \\
LENA\cite{LENA} & 7.4-9.4 & $3.37\pm0.06\pm0.23$ \\
CUSB\cite{CUSB} & 10.5 & $3.54\pm0.05\pm0.40$ \\
CLEO 83\cite{CLEO83} & 10.5 & $3.77\pm0.06\pm0.24$ \\
Crystal Ball\cite{XBAL} & 9.4 & $3.48\pm0.04\pm0.16$ \\
ARGUS\cite{ARGUS91} & 9.36 & $3.46\pm0.03\pm0.13$ \\
MD-1\cite{MD1-96} & 7.25-10.34 & $3.58\pm0.02\pm0.14$ \\ 
Previous Expts., Weighted Average & $\approx$9.5 & $3.58\pm0.07$ \\
CLEO 97 (this work) & 10.5 & $3.56\pm0.01\pm0.07$ \\
\end{tabular}
\caption{\label{tab:Xscts}Summary of inclusive cross section measurements.}
\end{table}

\section{Acknowledgments}
We gratefully acknowledge the effort of the CESR staff in providing us with
excellent luminosity and running conditions.
J.P.A., J.R.P., and I.P.J.S. thank                                           
the NYI program of the NSF, 
M.S. thanks the PFF program of the NSF,
G.E. thanks the Heisenberg Foundation, 
K.K.G., M.S., H.N.N., T.S., and H.Y. thank the
OJI program of DOE, 
J.R.P., K.H., M.S. and V.S. thank the A.P. Sloan Foundation,
R.W. thanks the 
Alexander von Humboldt Stiftung, 
and M.S. thanks Research Corporation
for support.
This work was supported by the National Science Foundation, the
U.S. Department of Energy, and the Natural Sciences and Engineering Research 
Council of Canada.


\begin{references}
\bibitem{ruski91}S. G. Gorishny, A. L. Kataev, S. A. Larin, Phys.
Lett. {\bf B259}, 144 (1991); L. R. Surguladze and M. A. Samuel,
Phys. Rev. Lett. {\bf 66}, 560 (1991), and K. G. Chetyrkin, 
Preprint hep-ph/9608480.
\bibitem{Teubner96}K. G. Chetyrkin, 
J. H. K\"uhn, and T. Teubner, Preprint TTP 96-35, 
DTP/96/80, hep-ph/9609411; K. G. Chetyrkin and J. H. K\"uhn, Phys. Lett.
{\bf B342}, 356 (1995); K. G. Chetyrkin and J. H. K\"uhn, Phys. Lett.
{\bf B308}, 127 (1993).
\bibitem{NIMLUM} CLEO Collaboration, G.~Crawford {\sl et al.},
Nucl.~Instrum.~Methods~Phys.~Res., Sect.~A {\bf 345}, 429 (1994).
\bibitem{r:CLEO-II} CLEO Collab., Y. Kubota $et~al$, 
                    Nucl. Instr. Meth. {\bf A320}, 66 (1992).
\bibitem{FoxWolf}G. C. Fox and S. Wolfram, Phys. Rev. Lett.
{\bf 41}, 1581 (1978), and G. C. Fox and S. Wolfram, Phys. Lett.
{\bf B82}, 134 (1979).
\bibitem{r:LUND} S. J. Sjostrand, LUND 7.3, CERN-TH-6488-92 (1992).

\bibitem{r:GEANT} R. Brun {\it et. al.}, GEANT v. 3.14, CERN Report No.
                  CERN CC/EE/84-1 (1987).
\bibitem{KORALB}We use KORALB (v.2.2) / TAUOLA (v.2.4).  References
for earlier versions are:
S. Jadach and Z. Was, Comput. Phys. Commun. {\bf 36}, 191 (1985);
{\bf 64}, 267 (1991); S. Jadach, J. H. K\"{u}hn, and Z. Was, Comput. Phys.
Commun. {\bf 64}, 275 (1991); {\bf 70}, 69 (1992); {\bf 76}, 361 (1993).
\bibitem{PDG96}Particle Data Group, R.M. Barnett, et al.,
Phys. Rev. {\bf D54}, (1996).
\bibitem{PLUTO}The PLUTO Collaboration, L. Criegee $et~al.$, 
Phys. Rep. C83 (1982), 151
\bibitem{DASPII}The DASP-II Collaboration, H. Albrecht $et~al.$, 
Phys. Lett. 116B (1982) 383
\bibitem{Heidelberg}P. Bock $et~al.$, Z. Phys. C 6 (1980) 125.
\bibitem{LENA}The LENA Collaboration, B. Niczporuck $et~al.$, 
Z. Phys. C 15, 299 (1982).
\bibitem{CUSB}The CUSB Collaboration, E. Rice $et~al.$, 
Phys. Rev. Lett. 48, 906 (1982).
\bibitem{CLEO83} The CLEO Collaboration, R. Giles $et~al.$, 
Phys. Rev. D29, 1285 (1984).
\bibitem{XBAL}The Crystal Ball Collaboration, Z. Jakubowski $et~al.$, 
Z. Phys. C 40, 49 (1988).
\bibitem{ARGUS91}The ARGUS Collaboration, H. Albrecht $et~al.$, 
Z. Phys. C 53, 13 (1991).
\bibitem{MD1-96}The MD-1 Collaboration, A.E. Blinov $et~al.$, 
Z.Phys. C 70, 31 (1996).
\end{references}
\end{document}